\documentclass[12pt]{myJHEP}
\usepackage[dvips]{graphicx}
\usepackage{latexsym}
\usepackage{cite}

\newcommand{\eqn}[1]{(\ref{#1})}
\newcommand{\be}{\begin{equation}}
\newcommand{\ee}{\end{equation}}
\newcommand{\no}{\nonumber}
\newcommand{\bel}[1]{\be\label{#1}}
\newcommand{\ba}{\begin{array}{c}}
\newcommand{\bat}{\begin{array}{cc}}
\newcommand{\bath}{\begin{array}{ccc}}
\newcommand{\ea}{\end{array}}
\newcommand{\beqn}{\begin{eqnarray}}
\newcommand{\eeqn}{\end{eqnarray}}

\newcommand{\bi}{\begin{itemize}}
\newcommand{\ei}{\end{itemize}}

\def\gap{\;\lower3pt\hbox{$\buildrel > \over \sim$}\;}
\def\lap{\;\lower3pt\hbox{$\buildrel < \over \sim$}\;}

\def\Comment#1{}

\newcommand{\Frac}[2]{\frac{\displaystyle #1}{\displaystyle #2}}

\title{$\mathbf{V_{\! us}}$ Determination from Hyperon Semileptonic Decays}

\author{V. Mateu and A. Pich \\ Departament de F\'\i sica Te\`orica, IFIC, Universitat de Val\`encia - CSIC
\\ Apt. Correus 22085, E-46071 Val\`encia, Spain
\\ E-mail: \email{Vicent.Mateu@ific.uv.es, Antonio.Pich@ific.uv.es}}

\abstract{We analyze the numerical determination of the quark mixing factor $V_{us}$ from
hyperon semileptonic decays. The discrepancies between the results obtained in two previous studies
are clarified. Our fits indicate sizeable SU(3) breaking corrections, which unfortunately can only be
fully determined from the data at the first order. The lack of a reliable theoretical calculation of
second-order symmetry breaking effects translates into a large systematic uncertainty, which has not
been taken into account previously. Our final result,
$|V_{us}| = 0.226\pm 0.005$, is not competitive with the existing determinations from
$K_{l3}$, $K_{l2}$ and $\tau$ decays.}


\preprint{IFIC/05$-$34\\ FTUV/05$-$0907 \\}

\begin{document}

\section{Introduction}

Accurate determinations of the quark mixing parameters
are of fundamental importance to test the flavour
structure of the Standard Model.
In particular, the unitarity of the CKM matrix
\cite{CA:63,KM:73} has been tested to the 0.2\% level
\cite{PI:05,CMS:04} with the precise measurement of its first-row
entries $|V_{ud}|$ and $|V_{us}|$ \cite{CKM:03}.
At that level of precision, a good control of systematic
uncertainties becomes mandatory. In fact, the existence of
small deviations from unitarity has been a long-standing
question for many years \cite{MS:86}.

Recently, there have been many relevant changes
to this unitarity test, which have motivated a very alive
discussion. While the standard $|V_{ud}|$ determination
from superallowed nuclear beta decays remains
stable, $|V_{ud}|= 0.9740\pm 0.0005$ \cite{CMS:04},
the information from neutron decay is
suffering strong fluctuations, due to conflicting data
on the axial coupling $g_A$ measured through decay
asymmetries \cite{PDG:04} and
the large decrease of the neutron
lifetime by more than $6\,\sigma$
obtained in the most recent precision measurement \cite{SE:05}.

On the other side, the $K\to\pi l\nu$ branching ratios have been
found to be significantly larger than the previously quoted world
averages. Taking into account the recently improved calculation of
radiative and isospin-breaking corrections
\cite{CNP:04,CKNRP:02,An:04}, the new experimental data from
BNL-E865 \cite{E865:03}, KTeV \cite{KTeV:04}, NA48 \cite{NA48:04}
and KLOE \cite{KLOE:04} imply \cite{Ci:05}
\begin{equation}
|V_{us}\, f_+^{K^0\pi^-}(0)| = 0.2175\pm 0.0005\, .
\end{equation}
In the SU(3) limit, vector current conservation guarantees that the
$K_{l3}$ form factor $f_+^{K^0\pi^-}(0)$ is equal to one. Moreover,
the Ademollo-Gatto theorem \cite{AG:64,Behrends:1960nf} states that
corrections to this result are at least of second order in SU(3)
breaking. They were calculated long-time ago, at $O(p^4)$ in Chiral
Perturbation Theory, by Leutwyler and Roos \cite{LR:84} with the
result $f_+^{K^0\pi^-}(0) = 0.961\pm 0.008$. Using the calculated
two-loop chiral corrections \cite{BT:03,PS:02}, two recent estimates
of the $O(p^6)$ contributions obtain the updated values
$f_+^{K^0\pi^-}(0) = 0.974\pm 0.011$ \cite{JOP:04} and
$f_+^{K^0\pi^-}(0) = 0.984\pm 0.012$ \cite{CEEKPP:05}, while a
lattice simulation in the quenched approximation gives the result
$f_+^{K^0\pi^-}(0) = 0.960\pm 0.005_{\mathrm{stat}} \pm
0.007_{\mathrm{syst}}$ \cite{BE:05} (the quoted lattice systematic
error does not account for quenching effects, which are
unfortunately unknown).
Taking $f_+^{K^0\pi^-}(0) = 0.974\pm 0.012$, one derives from $K_{l3}$:
\begin{equation}
|V_{us}| = 0.2233\pm 0.0028\, .
\end{equation}

An independent determination of $|V_{us}|$ can be obtained from the
Cabibbo-sup\-pressed hadronic decays of the $\tau$ lepton \cite{GJPPS:03}.
The present data implies \cite{GJPPS:04}
$|V_{us}| = 0.2208\pm 0.0034$.
The uncertainty is dominated by experimental errors in the 
$\tau$ decay
distribution and it is expected to be significantly improved at the B factories.
$|V_{us}|$ can be also determined from
$\Gamma(K^+\to\mu^+\nu_\mu)/\Gamma(\pi^+\to\mu^+\nu_\mu)$ \cite{MA:04},
using the lattice evaluation of the ratio of decay constants $f_K/f_\pi$ \cite{MILC:04};
one gets $|V_{us}| = 0.2219\pm 0.0025$.

The $|V_{us}|$ determination from hyperon decays is supposed to be affected
by larger theoretical uncertainties, because the axial-vector form factors
contributing to the relevant baryonic matrix elements are
not protected by the Ademollo-Gatto theorem. Thus, it suffers from first-order
SU(3) breaking corrections. Moreover, the second-order corrections to the
leading vector-current contribution are badly known. In spite of that, two
recent analyses of the hyperon decay data claim accuracies which,
surprisingly, are competitive with the previous determinations:
\begin{eqnarray}
|V_{us}| &=& 0.2250\pm 0.0027 \, , \qquad \mathrm{ref.} \mbox{\cite{CSW:03}} ,
\label{eq:CSW}\\ \label{eq:RFM}
|V_{us}| &=& 0.2199\pm 0.0026 \, , \qquad \mathrm{ref.} \mbox{\cite{RFM:04}} .
\end{eqnarray}
Although they use basically the same data, the two analyses result in
rather different central values for $|V_{us}|$ and obtain a qualitatively
different conclusion on the pattern of SU(3) violations. While
the fit of ref.~\cite{CSW:03} finds no indication of SU(3) breaking
effects in the data, ref.~\cite{RFM:04} claims sizeable second-order
symmetry breaking contributions which increase the vector form factors
over their SU(3) predictions. Clearly, systematic uncertainties seem
to be underestimated.

In order to clarify the situation, we have performed a new
numerical analysis of the semileptonic hyperon
decay data, trying to understand the differences between the results
\eqn{eq:CSW} and \eqn{eq:RFM}.
The theoretical description of the relevant decay amplitudes is briefly
summarized in section~\ref{sec:amplitudes}. The simplest phenomenological
fit in terms of the total decay rates and the experimental $g_1/f_1$ ratios
is presented in section~\ref{sec:g1f1}, which roughly reproduces
the numerical results of ref.~\cite{CSW:03}.
Section~\ref{sec:Nc} analyzes the sensitivity to SU(3) breaking, following the
same $1/N_C$ framework as refs.~\cite{RFM:04,FMJM:98}, while a discussion of systematic
uncertainties is given in section~\ref{sec:sys}.
For completeness, we discuss in section~\ref{sec:neutron}
the neutron-decay determination of $V_{ud}$. Our conclusions
are finally given in section~\ref{sec:summary}.

\section{Theoretical Description of Hyperon Semileptonic Decays}
\label{sec:amplitudes}

The semileptonic decay of a spin-$\frac{1}{2}$ hyperon,
$B_1\to B_2 \, l^-\bar\nu_l$,
involves the hadronic matrix elements of the vector and axial-vector currents:
\begin{eqnarray}
\langle B_2(p_2) |\, V^\mu\, | B_1(p_1)\rangle & = &
\bar u(p_2)\left[ f_1(q^2)\,\gamma^\mu + i\, {f_2(q^2)\over M_{B_1}}\,
\sigma^{\mu\nu} q_\nu + {f_3(q^2)\over M_{B_1}}\, q^\mu\right]
u(p_1)\, ,
\no\\ && \label{ffdef}\\
\langle B_2(p_2) |\, A^\mu\, | B_1(p_1)\rangle & = &
\bar u(p_2)\left[ g_1(q^2)\,\gamma^\mu + i\, {g_2(q^2)\over M_{B_1}}\,
\sigma^{\mu\nu} q_\nu + {g_3(q^2)\over M_{B_1}}\, q^\mu\right] \gamma_5\,
u(p_1)\, ,
\no\end{eqnarray}
where $q=p_1-p_2$ is the four-momentum transfer.
Since the corresponding $V-A$ leptonic current satisfies $q^\mu L_\mu \sim m_l$,
the contribution of the form factors $f_3(q^2)$ and $g_3(q^2)$
to the decay amplitude
is suppressed by the charged lepton mass $m_l$. Therefore,
these two form factors can be safely neglected in the electronic decays
which we are going to consider.

{\renewcommand{\arraystretch}{1.2}
\TABLE[thb]{
\begin{tabular}{|c||c|c|c|c|c|c|}\hline
$B_1\to B_2$ & $n\to p$ & $\Lambda\to p$ & $\Sigma^-\to n$ & $\Xi^-\to\Lambda$ &
$\Xi^-\to\Sigma^0$ & $\Xi^0\to\Sigma^+$
\\ \hline
$C_F^{B_2 B_1}$ & 1 & $-\sqrt{3/2}$ & $-1$ & $\sqrt{3/2}$ & $1/\sqrt{2}$ & $1$
\\ \hline
$C_D^{B_2 B_1}$ & 1 & $-1/\sqrt{6}$ & 1 & $-1/\sqrt{6}$ & $1/\sqrt{2}$ & $1$
\\ \hline
\end{tabular}
\caption{Clebsh-Gordan coefficients for octet baryon decays.}
\label{tab:CG}
}}

In the limit of exact SU(3) symmetry, the current matrix elements among the
different members of the baryon octet are related \cite{FMJM:98}:
\begin{eqnarray}
f_k^{\mathrm{sym}}(q^2) & = & C_F^{B_2 B_1}\, F_k(q^2) \, +\, C_D^{B_2 B_1}\, D_k(q^2) \, ,
\no\\ &&\label{cgf}\\
g_k^{\mathrm{sym}}(q^2) & = & C_F^{B_2 B_1}\, F_{k+3}(q^2) \, +\, C_D^{B_2 B_1}\, D_{k+3}(q^2) \, ,
\no\end{eqnarray}
where $F_k(q^2)$ and $D_k(q^2)$ are reduced form factors and
$C_F^{B_2 B_1}$ and $C_D^{B_2 B_1}$ are well-known Clebsh-Gordan coefficients.
The conservation of the vector current implies $F_3(q^2) = D_3(q^2)=0$.
Moreover, 
since the electromagnetic current belongs to the same octet of vector currents,
the values at $q^2=0$ of the vector form factors are determined by the electric charges and the
anomalous magnetic moments of the two nucleons,
$\mu_p = 1.792847351\, (28)$ and $\mu_n = -1.9130427\, (5)$ \cite{PDG:04}:
\begin{equation}\label{eq:magneticmoments}
F_1(0) = 1\, ,\qquad D_1(0) = 0\, ,\qquad
F_2(0) = -\left(\mu_p + \frac{1}{2}\,\mu_n\right)\, ,\qquad
D_2(0) = \frac{3}{2}\,\mu_n\, .
\end{equation}
The values at $q^2=0$ of the two reduced form factors determining $g_1(q^2)$
are the usual $F$ and $D$ parameters: $F_4(0) = F$, $D_4(0) = D$.
SU(3) symmetry also implies a vanishing ``weak-electricity'' form factor $g_2(q^2)$,
because charge conjugation does not allow a {\cal C}--odd $g_2$ term in the
matrix elements of the neutral axial-vector currents $A_\mu^3$ and  $A_\mu^8$,
which are {\cal C}--even.

The available kinematic phase space is bounded by
$m_e^2 \leq q^2 \leq (M_{B_1}-M_{B_2})^2$. Thus, $q^2$ is a parametrically small
SU(3) breaking effect. Since the form factor $f_2(q^2)$ appears multiplied
by a factor $q_\nu$, it gives a small contribution to the decay rate.
To $O(q^2)$ accuracy, which seems sufficient to analyze the current data,
the only momentum dependence which needs to be taken into account is the one of the
leading form factors $f_1(q^2)$ and $g_1(q^2)$:
\begin{equation}
\label{eq:qdep}
f_1(q^2) \approx f_1(0)\,\left( 1 + \lambda_1^f\, {q^2\over M_{B_1}^2}\right)\, ,
\qquad\qquad
g_1(q^2) \approx g_1(0)\,\left( 1 + \lambda_1^g\, {q^2\over M_{B_1}^2}\right)\, .\label{q2run}
\end{equation}
Moreover, $f_2(0)$ and $g_2(0)$ can be fixed to their SU(3) values, because any
deviations from the symmetry limit would give a second-order symmetry breaking effect.
Therefore, the form factor $g_2(q^2)$ can be neglected.

The slopes $\lambda_1^f$ and $\lambda_1^g$ are usually fixed assuming a
dipole form regulated by the mesonic resonance with the appropriate quantum
numbers \cite{GS:84,GK:85}: $f_1(q^2)= f_1(0)/(1 - q^2/M_V^2)^2$ and
$g_1(q^2)= g_1(0)/(1 - q^2/M_A^2)^2$; i.e.,
$\lambda_1^f = 2 M_{B_1}^2/M_V^2$ and $\lambda_1^g = 2 M_{B_1}^2/M_A^2$.
Previous analyses have adopted the mass values
$M_V = 0.97$~GeV \cite{CSW:03,RFM:04,FMJM:98,GS:84,GK:85} and
$M_A = 1.25$~GeV \cite{CSW:03,GS:84} or
$M_A = 1.11$~GeV \cite{RFM:04,FMJM:98,GK:85}.
We will
analyze in section~\ref{sec:sys} the systematic uncertainties associated
with these inputs.

It is useful to define the ratio of the physical value of $f_1(0)$
over the SU(3) prediction $C_F^{B_2 B_1}$:
\begin{equation}\label{ftilde}
\tilde{f}_1\, =\, f_1(0) / C_F^{B_2 B_1}\, =\,1+ \mathcal{O}(\epsilon^2) \, .
\end{equation}
Due to the Ademollo--Gatto theorem \cite{AG:64,Behrends:1960nf},
$\tilde{f}_1$ is equal to one up to second-order SU(3) breaking
effects.

The transition amplitudes for hyperon semileptonic decays have been
extensively studied, using standard techniques. We will not repeat
the detailed expressions of the different observables, which can be
found in refs.~\cite{RFM:04,GS:84,GK:85,Linke:1969aa}. In order to make a precision
determination of $|V_{us}|$, one needs to include the effect of radiative
corrections \cite{GK:85,SI:74,TSM:86,MTFMG:01}.
To the present level of experimental precision,
the measured angular correlation and angular spin-asymmetry coefficients
are unaffected by higher-order electroweak contributions. However, these corrections are
sizeable in the total decay rates. To a very good approximation,
their effect can be taken into account as a global correction to the
partial decay widths:
$\Gamma \sim G_F^2 |V_{us}|^2 (1 +\delta_{\mathrm{RC}})$.
The Fermi coupling measured in $\mu$ decay,
$G_F=1.16637\, (1)\cdot 10^{-5}\:\mathrm{GeV}^{-2}$
\cite{PDG:04},
absorbs some common radiative contributions. The numerical values of the
remaining corrections $\delta_{\mathrm{RC}}$ can be obtained from ref.~\cite{GK:85}.

\section{$\mathbf{g_1/f_1}$ Analysis}
\label{sec:g1f1}

\TABLE[th]{
\begin{tabular}{|c|c|c|c|c|c|}\hline
& $\Lambda\to p$ & $\Sigma^-\to n$ & $\Xi^-\to \Lambda$ &
$\Xi^-\to \Sigma^0$ & $\Xi^0\to \Sigma^+$
\\ \hline
$R$ & $3.161\pm 0.058$ & $6.88\pm 0.24$ & $3.44 \pm 0.19$ &
$0.53\pm 0.10$ & $0.93\pm 0.14$
\\
$\alpha_{e\nu}$ & $-0.019\pm 0.013$ & $0.347\pm 0.024$ & $0.53\pm 0.10$ &&
\\
$\alpha_e$ & $0.125\pm 0.066$ & $-0.519\pm 0.104$ &&&
\\
$\alpha_\nu$ & $0.821\pm 0.060$ & $-0.230\pm 0.061$ &&&
\\
$\alpha_B$ & $-0.508\pm 0.065$ & $0.509\pm 0.102$ &&&
\\
$A$ &&& $0.62\pm 0.10$ &&
\\ \hline
$g_1/f_1$ & $0.718\pm 0.015$ & $-0.340\pm 0.017$ & $0.25\pm 0.05$ &&
$1.32\pm 0.22$
\\ \hline
\end{tabular}
\caption{Experimental data on $|\Delta S| =1$ hyperon semileptonic decays \cite{PDG:04}.
$R$ is given in units of $10^6\:\mathrm{s}^{-1}$.}
\label{tab:measurements}
}

{\renewcommand{\arraystretch}{1.2}
\TABLE[th]{
\begin{tabular}{|c|c|c|c|c|}\hline
& $\Lambda\to p$ & $\Sigma^-\to n$ & $\Xi^-\to \Lambda$ &
$\Xi^0\to \Sigma^+$
\\ \hline
$|\tilde{f}_1\, V_{us}|$ &
$0.2221\, (33)$  
& $0.2274\, (49)$ & $0.2367\, (97)$ &  
$0.216\, (33)$  
%
%
%
\\ \hline
\end{tabular}
\caption{Results for $|\tilde{f}_1\, V_{us}|$ obtained from the
measured rates and $g_1(0)/f_1(0)$ ratios.
The quoted errors only reflect the statistical uncertainties.}
\label{tab:CabRes}
}}

The experimentally measured observables in hyperon semileptonic
decays \cite{PDG:04} are given in Table~\ref{tab:measurements},
which collects the total decay rate $R$, the angular correlation
coefficient $\alpha_{e\nu}$ and the angular-asymmetry coefficients
$\alpha_e$, $\alpha_\nu$, $\alpha_B$, $A$ and $B$. The precise
definition of these quantities can be found in refs.~\cite{GS:84,GK:85}.
Also given is the ratio $g_1(0)/f_1(0)$, which is determined from
the measured asymmetries.

The simplest way to analyze \cite{CSW:03} these experimental results
is to use the measured values of the rates and the ratios $g_1(0)/f_1(0)$.
Taking for $f_2(0)$ the SU(3) predictions, this determines
the product $|\tilde{f}_1\, V_{us}|$.
Table~\ref{tab:CabRes} shows the results obtained from the four
available decay modes. The differences with the values
given in ref.~\cite{CSW:03} are very small;
the largest one is due to the slightly different
value of the $\Xi^0\to\Sigma^+e^-\bar\nu_e$ branching ratio \cite{KTeV:99}.
The four decays give consistent results
($\chi^2/\mathrm{d.o.f.} = 2.52/3$),
which allows one (assuming a common value for $\tilde{f}_1$)
to derive a combined average
\begin{equation}\label{eq:CabRes}
|\tilde{f}_1\, V_{us}| \, = \, 0.2247 \pm 0.0026\, .
\end{equation}
This number agrees
(assuming $\tilde{f}_1=1$) with the value in Eq.~\eqn{eq:CSW}.

The quoted uncertainty only reflects the statistical errors and does not
account for the unknown SU(3) breaking contributions
to $\tilde{f}_1-1$, and other sources of theoretical uncertainties
such as the values of $f_2(0)$ and $g_2(0)$ [SU(3) has been
assumed], or the momentum dependence of $f_1(q^2)$ and $g_1(q^2)$.
We will estimate later on the size of all these effects.
For the moment, let us just mention that changing the dipole
ansatz for $f_1(q^2)$ and $g_1(q^2)$ to a monopole form,
the central value in \eqn{eq:CabRes} increases to
$0.2278$, with a $\chi^2/\mathrm{d.o.f.} = 3.24/3$.

The agreement among the four determinations in Table~\ref{tab:CabRes}
has been claimed to be a strong indication that SU(3) breaking
effects are indeed small \cite{CSW:03}. Note, however, that
first-order symmetry breaking corrections
in the ratio $g_1(0)/f_1(0)$ are effectively taken into account, since
we have used the experimental measurements.
What Table~\ref{tab:CabRes} shows is that the fitted results are
consistent, within errors, with a common $\tilde{f}_1$ value for the four
hyperon decays. The deviations of $\tilde{f}_1$
from one are of second order in symmetry breaking, but unfortunately
even their sign seems controversial \cite{DHK:87,Sch:95,Kr:90,AL:93}.

\section{$\mathbf{1/N_C}$ Analysis of SU(3) Breaking Effects}
\label{sec:Nc}

The limit of an infinite number of quark colours provides useful
simplifications of the strong interacting dynamics
\cite{HO:74,WI:79}, both in the meson \cite{PI:02,MA:98} and baryon
sectors \cite{MA:98}. The $1/N_C$ expansion of QCD provides a
framework to analyze the spin-flavour structure of baryons
\cite{DJM:94,DJM:95}, which can be used to investigate the size of
SU(3) breaking effects through a combined expansion in $1/N_C$ and
SU(3) symmetry breaking. A detailed analysis, within this framework,
of SU(3) breaking in hyperon semileptonic decays was performed in
ref.~\cite{FMJM:98,RFM:04}, where all relevant formulae can be
found. To avoid unnecessary repetition we will only show explicitly
the most important ingredients which have been used in the recent
$|V_{us}|$ determination of ref.~\cite{RFM:04}.

{\renewcommand{\arraystretch}{1.2} \TABLE[thb]{
\begin{tabular}{|c|c|c|}\hline
Decay & $g_1(0)$ & $\tilde{f}_1$
\\ \hline  
$n \to p$ & $\frac{5}{3}\,\tilde{a} + \tilde{b} + \rho$ & 1
\\
$\Lambda\to p$ & $-\sqrt{\frac{3}{2}}\,\left(\tilde{a} + \tilde{b} +
c_3 + c_4\right)$ & $1 + v_1 + v_2$
\\
$\Sigma^-\to n$ & $\frac{1}{3}\,\left(\tilde{a} + c_3 + c_4\right) -
\tilde{b}$ & $1 + v_1 + v_2 - 2 v_3$
\\
$\Xi^-\to \Lambda$ & $\frac{1}{\sqrt{6}}\,\left(\tilde{a} + 7\,
c_4\right) + \sqrt{\frac{3}{2}}\,\left(\tilde{b} + c_3\right)$ & $1
+ v_1 + 3\, v_2 + 2\, v_3$
\\
$\Xi^-\to \Sigma^0$ & $\frac{5}{3\sqrt{2}}\,\left(\tilde{a} + 3\,
c_3\right) + \frac{1}{\sqrt{2}}\,\left(\tilde{b} + c_4\right)$ & $1
+ v_1 + 3\, v_2$
\\
$\Xi^0\to \Sigma^+$ & $\frac{5}{3}\,\tilde{a} + \tilde{b} + 5\, c_3
+ c_4$ & $1 + v_1 + 3\, v_2$
\\ \hline
\end{tabular}
\caption{Parameterization of $g_1(0)$ and $\tilde{f}_1$ to first and
second order, respectively, in symmetry breaking \cite{FMJM:98}. The
neutron decay involves an additional parameter $\rho$, not included
in \eqn{eq:axial}.}
\label{tab:param} }}

At $q^2=0$ the hadronic matrix elements of the vector current are governed by the
associated charge or SU(3) generator. In the limit of exact SU(3) flavour symmetry,
$V^{0a}= T^a$ to all orders in the $1/N_C$ expansion,
where $T^a$ are the baryon flavour generators.
The SU(3) symmetry breaking corrections to $V^{0a}$ have been computed to second
order \cite{FMJM:98,JL:96}. For the hyperon $|\Delta S|=1$ decays that we are
considering, the final result can be written in the form \cite{FMJM:98}
\begin{equation}
V^{0a}\, =\, (1+ v_1)\, T^a + v_2\,\left\{ T^a, N_s\right\}
 + v_3\,\left\{ T^a, -I^2+J^2_s\right\}\, ,\label{secorderf}
\end{equation}
where $N_s$ counts the number of strange quarks, $I$ denotes the
isospin and $J_s$ the strange  quark spin. The parameters $v_i$
constitute a second-order effect in agreement with the
Ademollo-Gatto theorem \cite{AG:64,Behrends:1960nf}.

The $1/N_C$ expansion for the axial-vector current was studied in
refs.~\cite{DJM:95,DDJM:96}.
For the hyperon $|\Delta S|=1$ decay modes, one can write the result
in a simplified form which accounts for first-order symmetry breaking effects
\cite{FMJM:98,RFM:04}:
\begin{equation}\label{eq:axial}
{1\over 2}\,
A^{ia}\, =\, \tilde{a}\, G^{ia} + \tilde{b}\, J^i T^a +
c_3\,\left\{
G^{ia}, N_s\right\} + c_4\,\left\{ T^a, J^i_s\right\}\, .
\end{equation}
Here, $G^{ia} =
q^\dagger\left(\Frac{\sigma^i}{2}\otimes\Frac{\lambda^a}{2}\right)
q$ \ is a one-body quark operator acting on the spin and flavour
spaces. The coefficients $\tilde{a}\equiv a + c_1$ and
$\tilde{b}\equiv b + c_2$ reabsorb the effect of two additional
operators considered in ref.~\cite{FMJM:98}. These operators
generate an additional contribution to neutron decay, which we
parameterize as \ $\rho = - \left(\frac{5}{3}\, c_1 + c_2\right)$.
Table~\ref{tab:param} shows the resulting values of $g_1(0)$ and
$\tilde{f}_1$ for the relevant decay modes, in terms of the
parameters $\tilde{a}$, $\tilde{b}$, $c_3$, $c_4$, $\rho$, $v_1$,
$v_2$ and $v_3$.

\TABLE[thb]{
\begin{tabular}{|c|c|c|c|c|}\hline
& \multicolumn{2}{c|}{SU(3) symmetric fit} &
\multicolumn{2}{c|}{$1^{\mathrm{st}}$-order symmetry breaking}
\\ \hline
& Asymmetries & $g_1(0)/f_1(0)$ & Asymmetries & $g_1(0)/f_1(0)$
\\ \hline
$|V_{us}|$ & $0.2214\pm 0.0017$ & $0.2216\pm 0.0017$ &
$0.2266\pm 0.0027$ & $0.2239\pm 0.0027$
\\
$\tilde{a}$ & $0.805\pm 0.006$ & $0.810\pm 0.006$ &
$0.69\pm 0.03$ & $0.72\pm 0.03$
\\
$\tilde{b}$ & $-0.072\pm 0.010$ & $-0.081\pm 0.010$ &
$-0.071\pm 0.010$ & $-0.081\pm 0.011$
\\
$c_3$ &&& $0.026\pm 0.024$ & $0.022\pm 0.023$
\\
$c_4$ &&&  $0.047\pm 0.018$ & $0.049\pm 0.018$
\\ \hline
$\chi^2/\mathrm{d.o.f.}$ & $40.23/13$ &  $14.15/6$ & $18.09/11$ & $2.15/4$
\\ \hline
\end{tabular}
\caption{Results of different fits to the semileptonic hyperon decay data.}
\label{tab:fits}
}

In the strict SU(3) symmetry limit, $c_i = v_i = 0$, i.e.
$\tilde{f}_1=1$ while the values of $g_1(0)$ are determined by two parameters $a$ and $b$,
or equivalently by the more usual quantities $D=a$ and $F=\frac{2}{3}\, a + b$. A 3-parameter
fit to the hyperon decay data gives the results shown in Table~\ref{tab:fits}. Column~2
uses directly the measured values of the different rates and asymmetries, while in column~3
the asymmetries have been substituted by the derived $g_1(0)/f_1(0)$ values in
Table~\ref{tab:measurements}. Both procedures give consistent results, but the direct fit
to the asymmetries has a worse $\chi^2/\mathrm{d.o.f.} = 3.09$ (2.36 for the $g_1(0)/f_1(0)$
fit). These $\chi^2$ values indicate the need for SU(3) breaking corrections. The fitted
parameters agree within errors with the ones obtained in ref.~\cite{RFM:04}, although
our central value for $|V_{us}|$ is $1\,\sigma$ smaller. For the $F$ and $D$ parameters,
we obtain:
\begin{equation}
F = 0.462\pm 0.011 \, ,\qquad   D = 0.808\pm 0.006 \, ,\qquad
F+D = 1.270\pm 0.015\, .
\end{equation}
The last number, can be compared with the value of $g_1(0)/f_1(0)$ measured in neutron
decay: $\left[g_1(0)/f_1(0)\right]_{n\to p} = D+F = 1.2695\pm 0.0029$ \cite{PDG:04}.
Using the $|V_{ud}|$ value determined from superallowed nuclear beta decays and
the neutron lifetime quoted by the Particle Data Group \cite{PDG:04},
a more precise value, $\left[g_1(0)/f_1(0)\right]_{n\to p} = 1.2703\pm 0.0008$,
has been derived in ref.~\cite{CMS:04}.

Including first-order SU(3) breaking effects in $g_1(0)$, the fit has two more free
parameters. The fitted values are given in the last two columns of Table~\ref{tab:fits}.
The effect of SU(3) breaking manifests through a value of $\tilde{a}$ lower than
$a$ (i.e. $c_1=-0.10\pm 0.03\not= 0$, taking $a$ from the SU(3) fits), and the non-zero
value of $c_4$. The fit to the asymmetries has again a worse $\chi^2/\mathrm{d.o.f.} = 1.64$
than the $g_1(0)/f_1(0)$ fit ($\chi^2/\mathrm{d.o.f.} = 0.54$) and gives a $1\,\sigma$ higher
value of $|V_{us}|$. Taking $|V_{us}|$ from the best fit, its central value is about
$1\,\sigma$ higher than the value obtained with exact SU(3) symmetry. These results agree
within errors with the corresponding fits in ref.~\cite{RFM:04}.

One can repeat the fits including also the neutron decay, which
introduces the additional parameter $\rho$
. Taking $V_{ud}=
0.9740\pm 0.0005$ \cite{CMS:04}, this gives a sizeable measure of
SU(3) breaking, $\rho=0.16\pm 0.05$. The other parameters remain
unchanged.

\TABLE[thb]{
\begin{tabular}{|c|c|c|}\hline
& \multicolumn{2}{c|}{$2^{\mathrm{nd}}$-order symmetry breaking}
\\ \hline  & Asymmetries & $g_1(0)/f_1(0)$
\\ \hline
$|\left( 1 + v_1 + v_2\right)\, V_{us}|$ & $0.2280\pm 0.0034$ & $0.2220\pm 0.0038$
\\
$\tilde{a}$ & $0.69\pm 0.03$ & $0.74\pm 0.04$
\\
$\tilde{b}$ & $-0.075\pm 0.010$ & $-0.083\pm 0.011$
\\
$c_3$ & $0.03\pm 0.03$ & $0.02\pm 0.03$
\\
$c_4$ & $0.04\pm 0.02$ & $0.04\pm 0.02$
\\
$v_2$ & $0.01\pm 0.03$ & $0.04\pm 0.03$
\\
$v_3$ & $-0.004\pm 0.013$ & $-0.013\pm 0.014$
\\ \hline
$\chi^2/\mathrm{d.o.f.}$ & $16.5/9$ &  $0.53/2$
\\ \hline
\end{tabular}
\caption{Second order fits to the semileptonic hyperon decay data.}
\label{tab:secondOrder}
}

Ref.~\cite{RFM:04} presents the results of another fit, including second-order SU(3) breaking
effects in $\tilde{f}_1$ through the parameters $v_i$. The final value quoted for $|V_{us}|$
comes in fact from this fit, where $|V_{us}|$, $v_1$, $v_2$ and $v_3$ are fitted simultaneously
(together with $\tilde{a}$, $\tilde{b}$, $c_3$ and $c_4$), obtaining a very good
$\chi^2/\mathrm{d.o.f.} = 0.72/2 = 0.36$. We cannot understand the meaning of this
numerical exercise. While it is indeed possible to fit the data with the parameters
given in ref.~\cite{RFM:04}, one can obtain an infinite amount of different parameter sets
giving fits of acceptable quality, because there is a flat $\chi^2$ distribution in this case.
This can be easily understood looking to the last column in Table~\ref{tab:param}. From
the four analyzed $|\Delta S| =1$ hyperon semileptonic decays, one could only determine
the global factor $|V_{us} \left(1 + v_1 + v_2\right)|$, $v_2$ and $v_3$. It is
not possible to perform separate determinations of $|V_{us}|$ and $v_1$ because,
as shown in Eq.~\eqn{secorderf}, the contribution to the vector current of the flavour generator
$T^a$ is always multiplied by the same global factor $(1+v_1)$.

To assess the possible size of these second-order effects, we have also performed a
7-parameter fit to the data. The results are shown in Table~\ref{tab:secondOrder}.
Once more, the fit to the asymmetries has a worse
$\chi^2/\mathrm{d.o.f.}$ and gives a larger value for
$|V_{us} \left(1 + v_1 + v_2\right)|$.
The fitted values are consistent with the results in Table~\ref{tab:fits}
from the first-order fit.
Within the present experimental uncertainties, the 7-parameter fit is not able
to clearly identify any non-zero effect from second-order SU(3) breaking.
Notice, that in this numerical exercise one is only considering second-order
contributions to $\tilde{f_1}$, while $g_1(0)$ is still kept at first order.
Unfortunately, it is not possible at present to perform a complete second-order
analysis, owing to the large number of operators contributing to the
axial current at this order.

Comparing the results from all fits, it seems safe to conclude that the
$g_1(0)/f_1(0)$ ratios are less sensitive to SU(3) breaking than the asymmetries.
Therefore, we will take as our best estimate the corresponding first-order result
in Table~\ref{tab:fits},
\bel{eq:f1Vus}
|\tilde{f}_1\, V_{us}| \, = \, 0.2239 \pm 0.0027\, .
\end{equation}
This number is in good agreement with the simplest phenomenological fit
in Eq.~\eqn{eq:CabRes} and could give a very adequate estimate of $|V_{us}|$,
once the systematic uncertainties are properly included.

\section{Systematic Uncertainties}
\label{sec:sys}

In our analysis the lepton masses, and therefore the form factors $f_3(q^2)$ and $g_3(q^2)$,
have been neglected. This approximation does not introduce any relevant uncertainty
at the present level of experimental accuracy. The errors associated with radiative
corrections have been already taken into account in the fits,
together with the experimental uncertainties.
At first order in symmetry breaking, the main source of parametric uncertainties
comes from the numerical values of $f_2(0)$ and the slopes $\lambda_1^f$ and $\lambda_1^g$
governing the low-$q^2$ behaviour of the form factors $f_1(0)$ and $g_1(0)$.

Since the $f_2(q^2)$ contribution to the decay amplitude appears multiplied by $q_\nu$,
which is already a parametrically small SU(3) breaking effect, at $\mathcal{O}(\epsilon)$
the value of $f_2(0)$ can be fixed in the SU(3) limit from the proton and
neutron magnetic moments [see Eq.~\eqn{eq:magneticmoments}]. However, what appears
in the vector matrix elements~\eqn{ffdef} are the ratios $f_2(q^2)/M_{B_1}$. The SU(3) limit
can either be applied to $f_2(0)$ or $f_2(0)/M_{B_1}$, because the
baryon masses are the same for the whole octet multiplet in the limit of exact SU(3) symmetry.
Taking the physical baryon masses, the numerical results would be obviously different.
In order to estimate the associated uncertainty in $f_2(0)$ we will vary its value within
the range obtained with these two possibilities.

\TABLE[thb]{
\renewcommand{\arraystretch}{1.4}
\begin{tabular}[tbh]{|l|c|c|c|c|}
\hline
Parameter  & \multicolumn{2}{c|}{SU(3) symmetric fit} &
\multicolumn{2}{c|}{$1^{\mathrm{st}}$-order symmetry breaking}
\\ \hline
& Asymmetries & $g_1(0)/f_1(0)$ & Asymmetries & $g_1(0)/f_1(0)$
\\ \hline
$f_2^{\Lambda\rightarrow p} = 2.40\pm 0.20$ & ${}^{-0.0001}_{+0.0001}$ &
${}^{-0.0001}_{+0.0001}$ & ${}^{+0.0001}_{-0.0000}$ & ${}^{-0.0002}_{+0.0001}$
\\
$f_2^{\Sigma^-\rightarrow n} = -2.32\pm 0.28$ & ${}^{-0.0001}_{+0.0000}$ &
${}^{+0.0001}_{-0.0000}$ & ${}^{+0.0000}_{-0.0000}$ & ${}^{-0.0001}_{+0.0000}$
\\
$f_2^{\Xi^-\rightarrow \Lambda} = 0.178\pm 0.030$ & ${}^{+0.0000}_{-0.0000}$ &
${}^{+0.0000}_{-0.0000}$ & ${}^{+0.0000}_{-0.0000}$ & ${}^{+0.0000}_{-0.0000}$
\\
$f_2^{\Xi^-\rightarrow\Sigma^0} = -3.2\pm 0.6$ & ${}^{+0.0000}_{-0.0000}$ &
${}^{+0.0000}_{-0.0000}$ & ${}^{+0.0000}_{-0.0000}$ & ${}^{+0.0000}_{-0.0000}$
\\
$f_2^{\Xi^0\rightarrow\Sigma^+} = -4.4\pm 0.8$  & ${}^{+0.0000}_{-0.0000}$ &
${}^{+0.0000}_{-0.0000}$ & ${}^{+0.0000}_{-0.0000}$ & ${}^{+0.0000}_{-0.0000}$
\\
$M_A = 1.10\pm 0.09$ & ${}^{+0.0001}_{-0.0001}$ & ${}^{+0.0001}_{-0.0001}$ &
${}^{+0.0001}_{-0.0001}$ & ${}^{+0.0001}_{-0.0001}$
\\
$M_V = 0.91\pm 0.07$ & ${}^{+0.0005}_{-0.0006}$ & ${}^{+0.0004}_{-0.0005}$ &
${}^{+0.0002}_{-0.0002}$ & ${}^{+0.0005}_{-0.0006}$
\\ \hline
Total systematic error & 0.0006 & 0.0004 & 0.0002 & 0.0006
\\ \hline
\end{tabular}
\caption{Parametric uncertainties of the $V_{us}$ determination from hyperon decays}
\label{tab:sys}
}

The slopes $\lambda_1^f$ and $\lambda_1^g$ are determined from electroproduction and neutrino scattering data with nucleons, which are sensitive to the flavour-diagonal vector and axial-vector form factors
in the $Q^2=-q^2>0$ region. The obtained distributions are well fitted with dipole parameterizations
$G_{V,A}(Q^2)= G_{V,A}(0)/\left(1+Q^2/M_{V,A}^2\right)^2$,
with $M^0_V=(0.84\pm 0.04)~\mathrm{GeV}$ and $M^0_A=(1.08\pm 0.08)~\mathrm{GeV}$ \cite{GS:84}.
Extrapolating these functional forms to $q^2>0$, one gets a rough estimate of the needed hyperon
form factor slopes in the SU(3) limit. To account for SU(3) breaking, one usually modifies the parameters
$M_V$ and $M_A$ in a rather naive way, adopting the values
$M_V = M^0_V \left(m_{K^*}/m_{\rho}\right) =0.98\, \mathrm{GeV}$ and
$M_A = M^0_A \left(m_{K_1}/m_{a_1}\right) = 1.12\, \mathrm{GeV}$.
To estimate the systematic uncertainty associated with $\lambda_1^f$ and $\lambda_1^g$,
we adopt these dipole parameterizations, varying the values of the vector and axial-vector
mass parameters between $M^0_{V,A}$ and $M_{V,A}$.
As mentioned in section~\ref{sec:g1f1}, a monopole parameterization could lead to a significative
shift of the fitted $V_{us}$ value; however, in this case one should take different values
for the parameters $M_{V,A}$, in order to fit the $q^2<0$ data.

In Table~\ref{tab:sys} we show the sensitivity of the resulting $V_{us}$ value to these
parametric uncertainties. Columns 2 and 3 give the induced systematic errors in the 3-parameter
[SU(3) symmetric] fits, while columns 4 and 5 contain the corresponding numbers for the
5-parameter fits including first-order SU(3) breaking in $g_1(0)$. In both cases, we indicate
separately the estimates obtained for the fits to the asymmetries and the $g_1(0)/f_1(0)$ fits.
The numbers in the table show that the vector slope $\lambda_1^f$ is the dominant source
of parametric uncertainty. In any case, these uncertainties are much smaller than the statistical
errors of the corresponding fits.

\TABLE[th]{
\renewcommand{\arraystretch}{1.2}
\begin{tabular}[tbh]{|c|c|c|c|c|c|}
\hline
 Reference & $ \Lambda\rightarrow p $ & $ \Sigma^-\rightarrow n $ & $ \Xi^-\rightarrow \Lambda $ &
$\Xi^-\rightarrow \Sigma^0 $ &  $\Xi^0\rightarrow \Sigma^+$ \\ \hline
DHK'87 \cite{DHK:87} (quark model) & 0.987 & 0.987 & 0.987 & 0.987 & 0.987 \\
Sch'95 \cite{Sch:95} (quark model) & 0.976 & 0.975 & 0.976 & 0.976 & \\
Kr'90 \cite{Kr:90} (chiral loops) & 0.943 & 0.987 & 0.957 & 0.943 & \\
AL'93 \cite{AL:93} (chiral loops) & 1.024 & 1.100 & 1.059 & 1.011 & \\
\hline
\end{tabular}
\caption{Theoretical predictions for $\tilde{f_1}$.}\label{tab:f1}
}

At second order, one should take into account the unknown value of $g_2(0)$ and the
$\mathcal{O}(\epsilon^2)$ corrections to $f_1(0)$ and $g_1(0)$.
There exist a few estimates of $f_1(0)$ using quark models and baryon chiral lagrangians.
Unfortunately, they give rather different results as shown
in Table~\ref{tab:f1}. The quark-model calculations agree with the naive expectation that SU(3)
corrections should be negative, i.e. $\tilde{f_1}<1$ \cite{DHK:87,Sch:95}.
In contrast, the chiral-loop estimates obtain large corrections with opposite signs:
while ref.~\cite{AL:93} finds values for $\tilde{f_1}$ which are larger than one for all
analyzed decays, ref.~\cite{Kr:90} gets results more consistent with the quark-model
evaluations. The two references use slightly different chiral techniques, and are probably
taking into account different sets of Feynman diagram contributions. Clearly, a new and more
complete calculation is needed.

Nothing useful is known about $g_2(0)$ and the needed
$\mathcal{O}(\epsilon^2)$ corrections to $g_1(0)$. However, $g_2(0)$
is not expected to give a sizeable contribution, while
$g_1(0)/f_1(0)$ can be directly taken from experiment using the
phenomenological fit of section~\ref{sec:g1f1}. In fact, the
experimental $g_1(0)/f_1(0)$ ratios given in
Table~\ref{tab:measurements} assume already $g_2(0)=0$. Thus, the
value of $\tilde{f}_1$ constitute the main theoretical problem for
an accurate determination of $V_{us}$ from hyperon decays. Although
corrections to the SU(3) symmetric value are of
$\mathcal{O}(\epsilon^2)$, it has been argued that they are
numerically enhanced by infrared-sensitive denominators
\cite{AL:93,FMJM:98}. In the absence of a reliable theoretical
calculation, and in view of the estimates shown in
Table~\ref{tab:f1}, we adopt the common value
\bel{eq:f1_value} \tilde{f}_1\, =\, 0.99\pm0.02 \ee
for the five decay modes we have studied. While the two quark model
estimates are in the range $\tilde f_1 = 0.98\pm 0.01$, the
disagreement between the two chiral calculations expands the
interval of published results to $\tilde f_1 = 1.02\pm 0.08$.
However, for some decay modes such as $\Sigma^-\to n e^-\bar\nu_e$
one can show that $\tilde f_1$ should indeed be smaller than one, as
naively expected \cite{CSW:03,QB:68}. This disagrees with the
results obtained in ref.~\cite{AL:93}. Our educated guess in
\eqn{eq:f1_value} spans the whole interval of quark model results,
allowing also for higher values of $\tilde f_1$ within a reasonable
range.
Applying this correction to our best estimate in Eq.~\eqn{eq:f1Vus},
gives the final result:
\begin{equation}\label{eq:final}
|V_{us}|\, =\, 0.226\pm 0.005\, .
\end{equation}
%

\section{$\mathbf{V_{ud}}$ from Neutron Decay}
\label{sec:neutron}

A recent reanalysis of radiative corrections to the neutron decay
amplitude has given the updated relation
\cite{CMS:04,Garcia:2000gk}:
\begin{equation}
|V_{ud}|\, =\, \left( {4908\, (4)\:\mathrm{sec}\over \tau_n\,
\left( 1 + 3\, g_A^2\right)}\right)^{1/2}\, .\label{eq:nlt}
\end{equation}
Using $V_{ud}= 0.9740\pm 0.0005$, ref.~\cite{CMS:04} derives
the Standard Model prediction for the axial coupling
\begin{equation}
g_A \equiv g_1(0)/f_1(0) = 1.2703\pm 0.0008\, ,
\end{equation}
which is more precise than the direct measurements through neutron decay
asymmetries.

In order to extract $V_{ud}$ from \eqn{eq:nlt}, using as inputs the measured values
of the neutron lifetime and $g_A$, one would need to clarify the present
experimental situation. The Particle Data Group \cite{PDG:04}
quotes the world averages
\bel{eq:PDGvalues}
\tau_n = (885.7 \pm 0.8)~\mathrm{s}\, ,
\qquad\qquad
g_A = 1.2695\pm 0.0029\, ,
\ee
which imply
\bel{eq:VudPDG}
|V_{ud}|\, =\, 0.9745\pm 0.0019\, .
\ee
However, the most recent measurement of the neutron lifetime \cite{SE:05}
has lead to a very
precise value which is lower than the world average by $6.5\,\sigma$,
\bel{eq:Serebrov}
\tau_n = (878.5 \pm 0.7\pm 0.3)~\mathrm{s}\, .
\ee
Taking $g_A$ from \eqn{eq:PDGvalues}, this would imply a $2\,\sigma$ higher $|V_{ud}|$:
\bel{eq:Vud_Serebrov}
|V_{ud}|\, =\, 0.9785\pm 0.0019\, .
\ee

Actually, the PDG value of $g_A$ in \eqn{eq:PDGvalues} comes from an average of five
measurements which do not agree among them ($\chi^2 = 15.5$, confidence level = 0.004).
If one adopts the value obtained in the most recent and precise experiment \cite{AB:02},
\bel{eq:gA_Abele}
g_A = 1.2739\pm 0.0019\, ,
\ee
one gets the results:
\bel{eq:Vud_Abele}
|V_{ud}|\, =\, \left\{\:
\bat 0.9717 \pm 0.0013 & \qquad (\tau_n\;\mbox{from \cite{PDG:04}})\, ,
\\[10pt]
0.9757\pm 0.0013 & \qquad (\tau_n\;\mbox{from \cite{SE:05}})\, .\ea\right.
\ee

\section{Summary}
\label{sec:summary}

At present, the determinations of $|V_{ud}|$ and $|V_{us}|$ from baryon semileptonic decays
have large uncertainties and cannot compete with the more precise information obtained from
other sources. As shown in Eq.~\eqn{eq:nlt}, radiative corrections to the neutron decay
amplitude are known precisely enough to allow for an accurate measurement of $|V_{ud}|$, once
the existing experimental discrepancies will be resolved.
It looks surprising that two basic properties (lifetime and decay asymmetry)
of the neutron, one of the most stable particles, are still so badly known.
New precision experiments are urgently needed to clarify the situation.

Hyperon semileptonic decays could provide an independent determination of $|V_{us}|$, to be
compared with the ones obtained from kaon decays or from the Cabibbo-suppressed
$\tau$ decay width.
However, our theoretical understanding of SU(3) breaking effects constitutes a severe
limitation to the achievable precision.
We have presented a new numerical analysis of the available data, trying to understand
the discrepancies between the results previously obtained in
refs .~\cite{CSW:03} and ~\cite{RFM:04}, and the systematic uncertainties entering the
calculation.

\TABLE[t]{
\begin{tabular}{|c||c|c|c|c|}\hline
Source & $K_{l3}$ \
\cite{CNP:04,CKNRP:02,An:04,E865:03,KTeV:04,NA48:04,KLOE:04,LR:84,BT:03,PS:02,JOP:04,CEEKPP:05,BE:05}
& $K_{l2}$ \ \cite{MA:04,MILC:04} & $\tau$ \ \cite{GJPPS:03,GJPPS:04} & Hyperons
\\ \hline
$|V_{us}|$ & $0.2233\pm 0.0028$ & $0.2219\pm 0.0025$ & $0.2208\pm 0.0034$ &
$0.226\pm 0.005$
\\ \hline
\end{tabular}
\caption{Determinations of $V_{us}$.}
\label{tab:VusResults}
}

The $1/N_C$ expansion of QCD is a convenient theoretical framework
to study the baryon decay amplitudes and estimate the size of SU(3) breaking effects.
From the comparison of fits done at different orders in symmetry breaking, one
can clearly identify the presence of a sizeable SU(3) breaking at first order.
However, the present uncertainties are too large to pin down these effects at second order.

One can use the measured decay rates and $g_1(0)/f_1(0)$ ratios to
perform a rather clean determination of $|\tilde{f}_1\, V_{us}|$.
However it is impossible to disentangle $V_{us}$ from $\tilde{f}_1$
without additional theoretical input. The Ademollo-Gatto theorem
guarantees that $\tilde{f_1}=1 + \mathcal{O}(\epsilon^2)$, but it
has been argued that the second-order SU(3) corrections to
$\tilde{f_1}=1$ are numerically enhanced by infrared-sensitive
denominators \cite{AL:93,FMJM:98}. The existing calculations, using
quark models or baryon chiral perturbation theory, give
contradictory results and signal the possible presence of sizable
corrections. Adopting as an educated guess the value
$\tilde{f}_1=0.99\pm0.02$, we find our final result in
Eq.~\eqn{eq:final}.

Table~\ref{tab:VusResults} compares the hyperon determination of $V_{us}$, with
the results obtained from other sources. The present hyperon value has the largest
uncertainty. To get a competitive determination one would need
more precise experimental information and a better theoretical understanding
of $\tilde{f}_1$, beyond its symmetric value.
The average of all determinations is
\bel{eq:Final_Vus_average}
|V_{us}|\, =\, 0.2225\pm 0.0016\, .
\ee
Without the information from hyperon semileptonic decays, the average would be
$0.2221\pm 0.0016$. Taking $|V_{ud}|= 0.9740\pm 0.0005$,
from superallowed nuclear beta decays \cite{CMS:04}, the resulting first-row unitarity test
gives (the $|V_{ub}|$ contribution is negligible):
\bel{eq:unitarity}
|V_{ud}|^2 + |V_{us}|^2 + |V_{ub}|^2 = 0.9982 \pm 0.0012\, .
\ee
Thus, the unitarity of the quark mixing matrix is satisfied at the $1.5 \,\sigma$ level.

\acknowledgments{We have benefited from useful discussions with
Matthias Jamin, Rub\'en Flores--Mendieta, Aneesh V. Manohar and
Vicente Vento. This work has been supported in part by the EU
EURIDICE network (HPRN-CT2002-00311), the spanish Ministry of
Education and Science (grant FPA2004-00996), Generalitat Valenciana
(GRUPOS03/013) and by ERDF funds from the EU Commission.}


\end{document}